      [1999/12/01 v1.4c Il Nuovo Cimento]
\documentclass{cimento}

\usepackage{epsfig}

\title{Spectral Analysis of GRBs Measured by RHESSI}
\author{C.~Wigger\from{ins:psi}\ETC,
W.~Hajdas\from{ins:psi},
A.~Zehnder\from{ins:psi},
K.~Hurley\from{ins:ucb},
E.~Bellm\from{ins:ucb},
S.~Boggs\from{ins:ucb},
M.~Bandstra\from{ins:ucb},
        \atque
D.M.~Smith\from{ins:scipp}}
\instlist{
 \inst{ins:psi} Paul Scherrer Institute - Villigen PSI, Switzerland
 \inst{ins:ucb} Space Sciences Laboratory, University of California -
    Berkeley, CA, USA
 \inst{ins:scipp} Departement of Physics, University of California -
    Santa Cruz, CA, USA
}
\PACSes{
\PACSit{95.55.Ka}{X- and $\gamma$-ray telescopes and instrumentation}
\PACSit{95.55.Qf}{Photometric, polarimetric, and spectroscopic instrumentation}
\PACSit{98.70.Rz}{$\gamma$-ray sources; $\gamma$-ray bursts}
\PACSit{01.30.Cc}{Conference proceedings}
}
\begin{document}

\maketitle

\begin{abstract}

The Ge spectrometer of the RHESSI satellite
is sensitive to Gamma Ray Bursts (GRBs)
from about $40\,$keV up to $17\,$MeV, thus
ideally complementing the Swift/BAT instrument whose
sensitivity decreases above $150\,$keV.
We present preliminary results of spectral fits 
of RHESSI GRB data.
After describing our method, the RHESSI results are 
discussed and compared with Swift and Konus.

\end{abstract}

\section{Introduction}
%------------------------
The energy spectra of Gamma Ray Bursts (GRBs) are an important
element for a complete understanding of the GRB phenomenon. 
Most GRB spectra peak at some energy $E_{peak}$
in a $\frac{dN}{dE} E^2$ representation.
For GRBs with a measured redshift, the distance
can be estimated, and, combined with 
the observed fluence,
the total isotropic energy $E_{iso}$ can be determined.
In several publications, Amati et al.\ (\cite{ref:amati02} and
e.g.~\cite{ref:amati_c} and references therein) 
show that $E_{iso}$ and $E_{peak}$ are strongly 
correlated.
Correcting $E_{iso}$ for beaming, to derive the intrinsic
energy $E_\gamma$, Ghirlanda et al.\ 
(\cite{ref:ghirlanda2004}, \cite{ref:ghirlanda2005})
find a strong $E_{peak}$-$E_\gamma$ correlation.
It has been argued that both relations might be an artefact
of selection effects (\cite{ref:NP2005},\cite{ref:BP2005}).
However, it has been shown by
Ghirlanda et al.\ (\cite{ref:GGF2005}) that BATSE GRBs with
known redhift are consistent with the $E_{peak}$-$E_{iso}$
correlation.

A larger sample of GRBs may hold the key to understanding
spectral-energy correlations in general.
Since Swift's launch, the number of GRBs with measured
redshift has increased rapidly. However, the Swift BAT
energy sensitivity -- steeply declining above about $150\,$keV --
does not allow the determination of the
peak energy for many GRBs.
In order to fully determine peak energy and fluence,
information from other spacecraft with better
sensitivity at higher energies is needed.

\section{Instrument}
%---------------------
The Reuven Ramaty High Energy Solar Spectroscopic Imager
(RHESSI, \cite{ref:Lin2002})
is a NASA small explorer mission and was designed to study
solar flares in hard X-rays and $\gamma$-rays.
It was launched in 2002 into a low Earth orbit
(580 kilometer altitude, 38 degree inclination).
   RHESSI consists of two main parts:
   the imaging telescope and 
   the spectrometer behind it.
The RHESSI spectrometer 
(\cite{ref:ds2002}) consists of 9 %coaxial 
germanium detectors,
each $7.1\,$cm in diameter and $8.5\,$cm high. They are segmented
into a thin front ($\approx 1.5\,$cm) and a thick rear segment 
($\approx 7\,$cm).
Each detected photon is time- and energy-tagged from
$3\,$keV to $2.8\,$MeV (front) or $20\,$keV to $17\,$MeV (rear).
The energy resolution is $\approx 3\,$keV at $1\,$MeV, and the
time resolution is 1$\,\mu$s.
RHESSI always points towards the Sun and rotates
about its axis at $15\,$rpm. 

Since the shielding of the rear segments is minimal,
photons with more than about $25\,$keV can enter
from the side.
Above about 50 -- 80 keV, photons from any
direction can be detected.
Thus, for GRBs, RHESSI views 65\% of the sky, the rest 
being occulted by the Earth.
The effective area for GRB detection  
depends on the incident photon energy %$E$ 
and the angle between the GRB direction
and the RHESSI axis ('polar angle').
Over a wide range of energies and polar angles, the
effective area is around 150$\,$cm$^2$.
The sensitivity drops rapidly at energies below 
$\approx 50\,$keV. 
   %For sources with small polar angles (i.e.\ close to the Sun),
   %the detectors are partly shielded by the rest of the spacecraft, 
   %thus the sensitivity is reduced below $100\,$keV.
RHESSI observes 1--2 GRBs per week.

\section{Method}
%----------------
The response of RHESSI to $\gamma$-rays is
simulated using GEANT3, the simulation software for
high energy physics experiments by CERN. % \cite{ref:Wojtek??}.
Knowing the burst direction, we simulate
$\gamma$-rays with the corresponding polar angle
and a simple power law spectrum, i.e. $(\frac{dN}{dE})_{sim}
\propto E^{-\alpha}$ with typically $\alpha = 2\,$.
Rotation angles are generated with uniform distribution, 
i.e.~ the response function is spin averaged. 
The output of the simulation is
an event list
  containing the deposited energy as well as
  the initial photon energy.

The observed GRB spectrum -- in the form of a histogram --
is compared with a spectrum accumulated from the simulated
event list. 
In order to generate an arbitrary spectral shape, 
we apply a weighting factor to each
simulated event in the list. 
These factors would all be 1,
if the GRB spectrum had the same form
as the simulated one.
In general, the weighting factors are a 
function of the initial energy and the parameters
of the spectral shape. 

The most often used spectral shape is
the Band function (\cite{ref:band93}), 
which is a smooth combination of a
cut off power law at lower energies 
($ dN \propto E^{\alpha} \, \exp(-E/E_0) \,dE $) 
and a decaying simple power law at higher energies
($ dN \propto E^{\beta} \,dE$).
If $\beta < -2$ and $\alpha > -2$ then $E_{peak} = E_0(2+\alpha)\,$.

\section{Examples of fitted GRB spectra and ongoing work}
%----------------------------------------------

\begin{figure}[htbp]
  \centering
  \epsfig{file=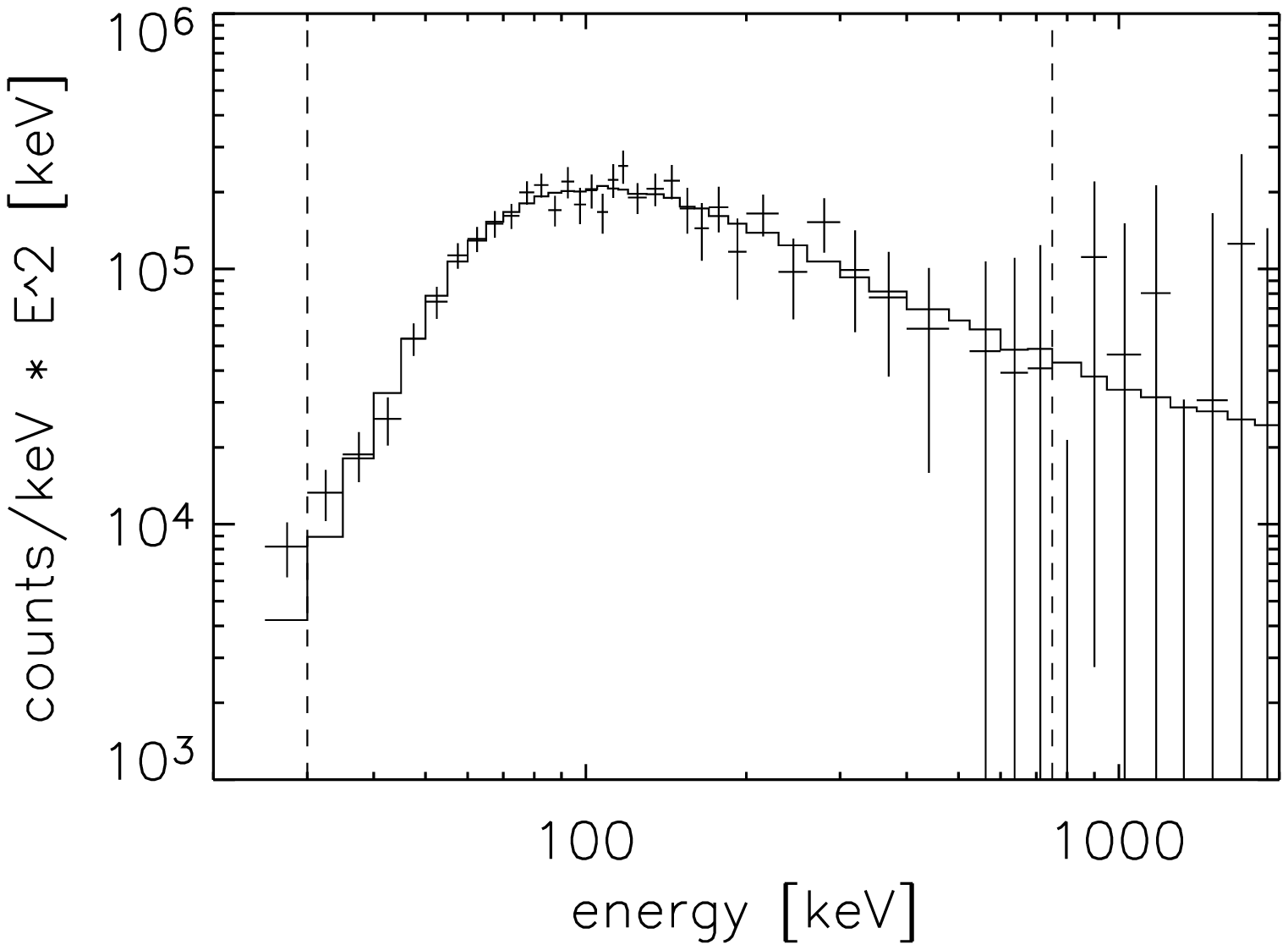,width=6.1cm} 
  \hfill
  \epsfig{file=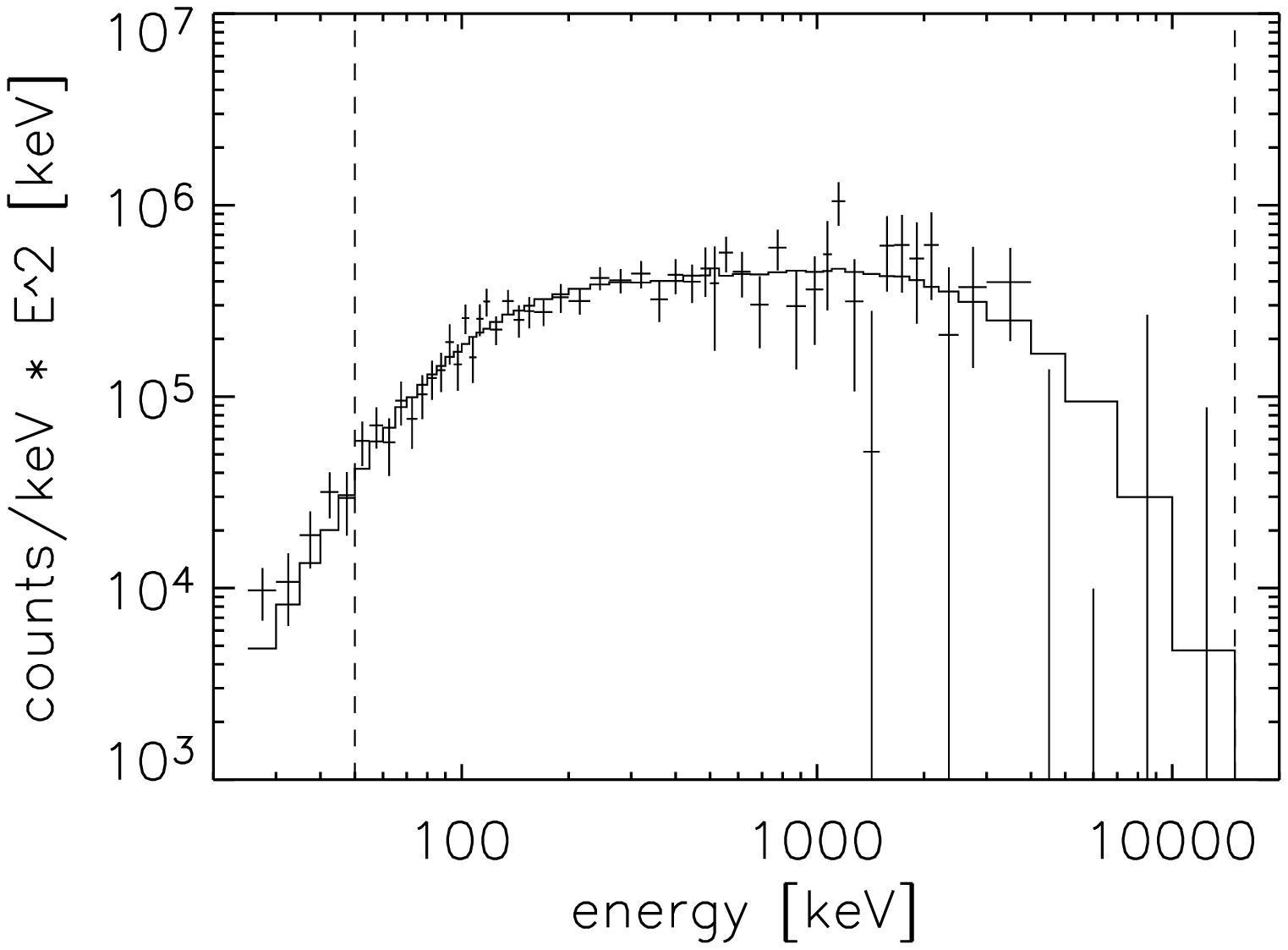,width=6.1cm}
  \caption{Spectra of GRB 050525A (left) and GRB 050717 (right).
    The RHESSI data are shown as crosses - with 1 sigma error bars -,
    the fit as line histogram,
    and the fit range is indicated by the
    dashed vertical lines.
  \label{fig:spectra}}
\end{figure}

The spectral fits of
GRB 050525A \cite{GCN050525A} and GRB 050717 \cite{GCN050717} 
are shown in fig.~\ref{fig:spectra},
and the RHESSI fit parameters are summarized and compared with 
the Swift and Konus fit parameters 
in table  \ref{tab:specfit}.
   For energies above the Swift/BAT energy range, 
the RHESSI results for the spectral parameters
(typically $\beta$ and $E_{peak}$) are well
determined and better constrained than the Konus values. 
In the lower energy range, the RHESSI spectral parameters
still agree within errors with the Konus and Swift/BAT results.

\begin{table}[htbp]
  \caption{Spectral parameters of GRB 050525A and GRB 050717}
  \label{tab:specfit}
  \begin{tabular}{llcccc} \hline
   Instrument & Ref. & $\alpha$
             & $E_{peak}$ (keV)
             & $\beta$
             & $f$  (erg/cm$^2$) \\
            \hline
   \multicolumn{6}{c}{\em GRB 050525A } \\
    \hline
  RHESSI  & \rule{0mm}{3mm}
              & 0.8 $^{+1.8}_{-3.0}$
              & 75 $^{+14}_{-10}$
              & $-$2.62 $^{+0.15}_{-0.24}$
              & (1.43 $^{+0.17}_{-0.12}$)$\cdot 10^{-5}$ $\;^{(1)}$\\
   Swift/BAT  &  \cite{ref:bl06}  \rule{0mm}{4mm}
              & $-$0.99 $^{+0.11}_{-0.12}$
              &  78.8 $^{+3.9}_{-3.1}$
              & $-\infty$ 
	      &\parbox{10em}{($2.01 \pm 0.05 $)$\cdot 10^{-5}$ $\;^{(1)}$} \\
   Konus      &  \cite{gcn_konus_050525}   \rule{0mm}{3mm}
              & $-$1.10 $\pm 0.05 $
              &  84.1 $\pm 1.7$
              & $-\infty$ & --   \\
   \hline	      
   \multicolumn{6}{c}{\em GRB 050717 }  \\
   \hline
  RHESSI &  %\rule{0mm}{3mm}
              & $-$1.14 $^{+0.17}_{-0.15}$   
              & 1550 $^{+510}_{-370}$   
              & $-\infty$
              & \parbox{10em}{($1.25 \pm 0.07$)$\cdot 10^{-5}$  $\;^{(1)}$ 
                  \hspace*{0.34em} ($5.4 \pm 0.5$)$\cdot 10^{-5}$  $\;^{(2)}$ } \\
   Swift/BAT  & \cite{ref:krimm}  \rule{0mm}{3.2mm}
              & $-1.0$ to $-1.5$ $^{(3)}$
              & $\infty$ 
              & $-\infty$ & ($1.40 \pm 0.03$)$\cdot 10^{-5}$  $\;^{(1)}$\\
   Konus &  \cite{ref:krimm}   \rule{0mm}{3.3mm}
              & $-$1.19 $\pm 0.12 $
              & 2101 $^{+1934}_{-830}$   
              & $-\infty$ 
	      & (6.5 $^{+0.9}_{-2.2}$)$\cdot 10^{-5}$ $\;^{(2)}$ \\
\hline	      
  \end{tabular}
  \footnotesize{The RHESSI errors and and the errors of Refs. 
  \cite{ref:bl06} and \cite{ref:krimm} are given with 90\% C.L.} \\
  \footnotesize{$^{(1)}$ 15-350 keV} \hspace{3.5em}
  \footnotesize{$^{(2)}$ 20-6000 keV} \hspace{3.5em}
  \footnotesize{$^{(3)}$ time dependent values (softening)}   
\end{table}

\vspace{1ex}
%\section{Ongoing work}
%---------------------

\begin{figure}[htbp]
  \centering
  \epsfig{file=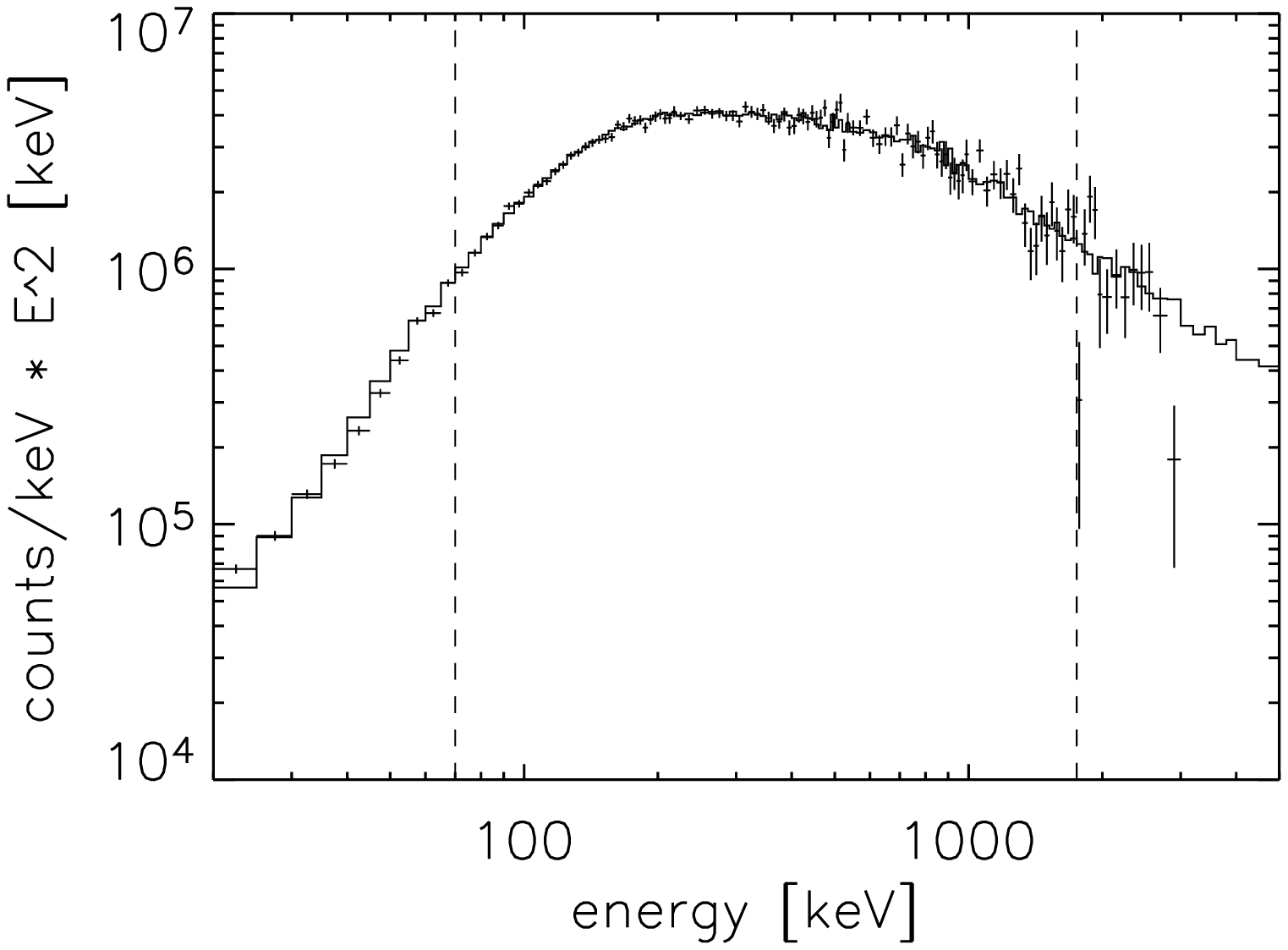,width=6.1cm} 
  \hfill
  \epsfig{file=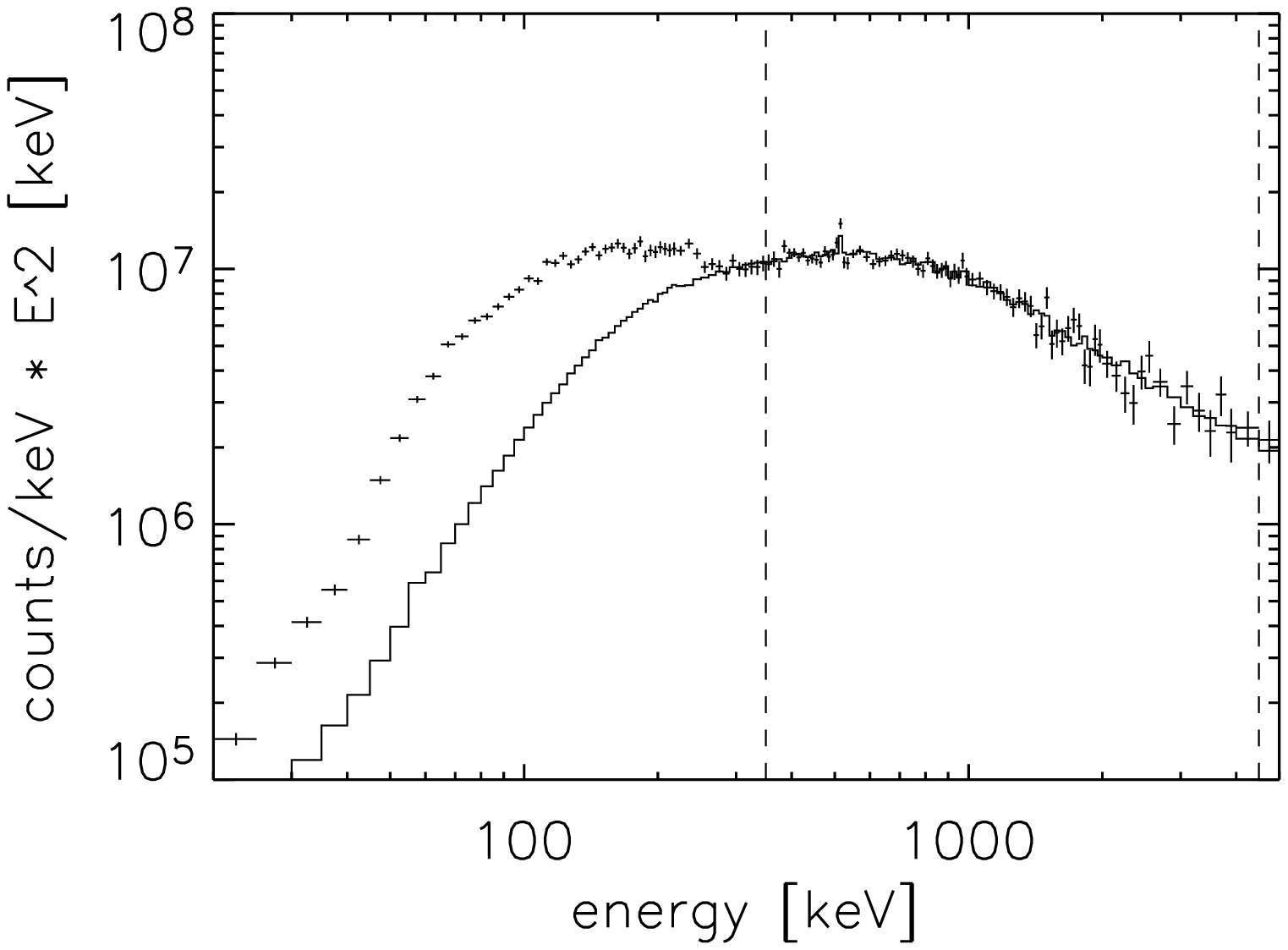,width=6.1cm}
  \caption{Spectrum of GRB 021206. 
    The front spectrum (left) and 
    the rear spectrum (right) were fitted independently
    with agreeing results.
    The fit ranges are indicated by the vertical dashed lines. 
    The weighted means of the spectral parameters are ($1 \sigma$ errors):
    $\alpha =  -0.65 \pm 0.04$,
    $E_{peak} = 705 \pm 14 \,$keV, and
    $\beta = -3.19 \pm 0.06$.
    The excess events observed in the rear segments (right plot)
    are due to GRB photons backscattered from the atmosphere. 
  \label{fig:grb021206}}
\end{figure}

The spectral fit 
of another GRB, namely GRB 021206, 
is presented in fig.~\ref{fig:grb021206}.
This GRB is famous for its debated polarization
(\cite{ref:CB03},\cite{ref:PSI04},\cite{ref:RF04},\cite{ref:roma04}).

The huge number of excess events observed in the rear
detectors below $300\,$keV is due to GRB photons that 
were backscattered from the atmosphere.
The geometrical constellation of the GRB, RHESSI, and Earth
was such that the GRB photons came from the front
direction, where the effective area is relatively small,
whereas the Earth was behind RHESSI so that the 
backscattered photons could easily reach the rear
segments. %-- We are working on simulating this effect.
In order to simulate this effect, 
we have constructed an atmospheric scattering model.
The first preliminary results show
qualitative agreement with the observed spectrum.

\section{Summary, Conlusion and Outlook}
%---------------------------------------

Even though RHESSI was designed to observe solar flares,
it is also a capable GRB detector with an
energy range from about 40 -- 80$\;$keV 
(depending on the direction of the GRB) up to 17$\;$MeV. 
With its wide field of view and its
excellent energy resolution, RHESSI is a useful 
high-energy complement to Swift/BAT
for measuring GRB spectra.
About one fourth of the Swift GRBs are also detected by RHESSI.
We have developed tools for spectral analysis.  
Our spectral parameters agree within uncertainties with those obtained
from other instruments. The RHESSI errors on the high energy parameters
tend to be smaller compared with other intstruments.
First GCN messages with spectral parameters by RHESSI
were published (\cite{gcn:hessi},\cite{gcn:hessiII},\cite{gcn:hessiIII}).

%\acknowledgments

%****
\end{document}